\documentclass[review]{elsarticle}

\usepackage{amsmath}
\usepackage[T1]{fontenc}
\usepackage{changepage}
\usepackage{graphicx}
\usepackage[hidelinks]{hyperref}
\usepackage{lmodern}
\usepackage{multirow}
\usepackage{subcaption}
\usepackage{algorithmic}
\usepackage{algorithm}
\usepackage{paralist}
\usepackage[disable]{todonotes}

\usepackage{newfloat}
\DeclareFloatingEnvironment[name=Listing]{listingfloat}

\usepackage{listings}
\graphicspath{{./figures/}}
\DeclareGraphicsExtensions{.pdf}
\journal{Computer Languages, Systems and Structures}
\usepackage[absolute]{textpos}
\makeatletter
\def\ps@pprintTitle{
\def\@oddfoot{}
}
\makeatother

\begin{document}

\begin{textblock}{15}(0.5,14.5)
{
\vspace{-2pt}\noindent\hrulefill

\noindent\fontsize{8pt}{8pt}\selectfont\copyright\ 2017. This manuscript version is made available under the CC-BY-NC-ND 4.0 license: \url{http://creativecommons.org/licenses/by-nc-nd/4.0/}. \hspace{5pt} This is the accepted version of: M. Nos\'a\v{l}, J. Porub\"an, M. Sul\'ir. Customizing host IDE for non-programming users of pure embedded DSLs: A case study. Computer Languages, Systems and Structures (COMLAN), Vol. 49, 2017, pp. 101--118. \url{http://doi.org/10.1016/j.cl.2017.04.003}

}
\end{textblock}

\begin{frontmatter}

\title{Customizing Host IDE for Non-programming Users of Pure Embedded DSLs: A Case Study}

\author[svagant]{Milan Nos\'a\v{l}\corref{cor1}}
\ead{milan.nosal@gmail.com}
\cortext[cor1]{Corresponding author}
\author[tuke]{Jaroslav Porub\"an}
\ead{jaroslav.poruban@tuke.sk}
\author[tuke]{Mat\'u\v{s} Sul\'ir}
\ead{matus.sulir@tuke.sk}
\address[svagant]{Independent / Svagant\\
Mlynsk\'{a} 28, 040 01 Ko\v{s}ice, Slovakia}
\address[tuke]{Department of Computers and Informatics\\
Faculty of Electrical Engineering and Informatics\\
Technical University of Ko\v{s}ice\\
Letn\'a 9, 042 00 Ko\v{s}ice, Slovakia}

\begin{abstract}
Pure embedding as an implementation strategy of domain-specific languages (DSLs) benefits from low implementation costs. On the other hand, it introduces undesired syntactic noise that impedes involvement of non-programming domain experts. Due to this, pure embedded DSLs are generally not intended for, nor used by, non-programmers. In this work, we try to challenge this state by experimenting with inexpensive customizations of the host IDE (Integrated Development Environment) to reduce the negative impact of syntactic noise. We present several techniques and recommendations based on standard IDE features (e.g., file templates, code folding, etc.) that aim to reduce syntactic noise and generally improve the user experience with pure embedded DSLs. The techniques are presented using a NetBeans IDE case study. The goal of the proposed techniques is to improve the user experience with pure embedded DSLs with a focus on the involvement of non-programming domain experts (or non-programmers in general). The proposed techniques were evaluated using a controlled experiment. The experiment compared a group using Ruby and non-modified RubyMine IDE versus a group using Java and NetBeans IDE customized to use the proposed techniques. Experiment results indicate that even inexpensive host IDE customizations can significantly alleviate issues caused by the syntactic noise: Java with its inflexible syntax performed better than Ruby with its concise syntax.
\end{abstract}


\end{frontmatter}

\section{Introduction}
\label{sec:introduction}

Domain-specific languages (DSLs) are languages tailored for a specific, rather narrow domain \cite{preliminaryStudy,whenAndHow}. The main advantages of DSLs are high abstraction level, high expressivity due to usage of domain concepts in a language, and involvement of domain experts \cite{whenAndHow,dslVanDeursen,designPatternsDSL,hudak1997domain}. As a result, applications of the DSL approach can be found in many domains, such as secure logging \cite{mernikFAL}, questionnaires \cite{questionnaireDSL}, database design \cite{sonjaConstraint}, feature modeling \cite{vranicOrgPatterns}, expert systems \cite{polap}, etc. Disadvantages are mainly centered around language creation costs and insufficient tool support \cite{whenAndHow,extensibleLanguage}.

From the practical point of view, there are two common possibilities of DSL implementation\footnote{For a more thorough discussion of the categorization of DSL approaches and the terminology, we refer the reader to \cite{mernik2013}.}: embedded DSLs (EDSLs) and stand-alone (external) DSLs.

A stand-alone DSL has its own parser and thus does not pose any restrictions on the DSL syntax. However, the creation cost of such a language can be high. Not only is it quite time-consuming, but it also poses high mental and knowledge requirements on the developer, like proficiency in context-free grammars and specialized tools (e.g., parser generators, language workbenches, etc.) \cite{infoQ,fowlerDSL,erikM}.

From the implementation viewpoint, there exist heterogeneous and homogeneous embedded DSLs \cite{mernik2013}. Heterogeneous embedding utilizes language composition on the level of tools -- in heterogeneous embedding, the embedded language and the host language are not processed by the same compiler. An example of heterogeneous embedding are island grammars \cite{boundedSeas}. ``Islands'' are grammars embedded in the host language grammar, which is called ``water''. ``Islands'' are parsed using their own parser, the ``water'' is parsed by the host language parser. 

Homogeneous embedding, on the other hand, reuses a host language compiler to reduce implementation costs \cite{mernik2013}. As an example of such embedding, we can mention extensible compilers -- an extended compiler parses both the embedded and host language. A special case of homogeneous embedding is pure embedding (first coined by Hudak \cite{hudak}) when the host compiler is fully reused: the pure EDSL is a valid sentence in the host language, without a need of preprocessing or compiler extension. In the case of pure embedding, the DSL uses the host language's syntax to embed a DSL into a host GPL (general-purpose language). Basically, a pure EDSL is an API (application programming interface) designed with a strong focus on fluency, readability and the use of domain-specific terms. For its low creation costs, pure EDSLs are an obvious choice for many real-life applications and for DSL prototyping.

Faster creation of pure EDSLs comes at a price: they contain ``syntactic noise'' like brackets or import statements. Therefore, pure embedding poses an important question to the developer: What host language to choose? Some dynamically typed languages with very flexible syntax, e.g., Ruby, offer a low level of syntactic noise \cite{cunningham,embeddingRuby} -- there is only a small portion of syntax elements unrelated to the problem domain. This is the reason they are a common choice for the EDSL approach. However, large codebases are written in languages with rigid syntax like Java, which is also used to implement pure EDSLs \cite{fowlerDSL,freeman}, although less commonly. Furthermore, there is a much larger community of programmers available (i.e., a larger workforce). 

The main drawback of syntactic noise is that it impedes non-programming domain experts' involvement. For this reason, one could come to a conclusion that an EDSL is just a language with a higher level of abstraction meant only for programmers. One of the characteristics of DSL programs is that they can often be written by non-programming domain experts \cite{hudak1997domain}. However, pure EDSLs, or EDSLs in general, are often designed only for programmers. We try to challenge this state by using small, inexpensive IDE customizations to support EDSL usage by non-programmers. Our work is based on the idea that when we evaluate the design of DSLs, we should consider \emph{not only their textual expressiveness but also IDE support}. Therefore, the goal of this paper is twofold.

First, \emph{we will show how an existing GPL IDE can be used to facilitate usage of pure EDSLs by non-programmers}. Although in our work we focus on \mbox{non-programmers}, programmers can benefit from these techniques as well. We will present a set of techniques achieved by small and inexpensive IDE customizations, requiring almost no effort to implement.

Second, \emph{we aim to show how proper IDE support for a syntactically rigid language like Java can compete with, or even outweigh the advantages of a flexible language like Ruby} (Ruby, along with Groovy or Clojure, is one of the most preferable languages for pure embedding \cite{ghosh2011dsl}). In the DSL community there is a lack of evaluation research which was outlined also in other studies \cite{DSLSystematicMappingStudy,SMSKosar}. Therefore, we performed a controlled experiment with non-programmers, comparing a Java group with small IDE customizations vs. a Ruby group with a standard IDE.

Contributions of this work can be stated as follows:

\begin{itemize}
	\item a description of techniques aiming to improve the experience of non-programmers (and also programmers) with writing pure EDSL programs,
	\item implementation of a NetBeans plugin for folding and guarding surrounding noise for the purposes of a case study,
	\item minor implementation of origin tracking for domain error messages that piggybacks on NetBeans IDE, and
	\item experimental evaluation of the aforementioned techniques on non-programmers.
\end{itemize}

The case study presented in this paper is performed on a combination of the Java language and the NetBeans IDE. We used Java, since due to its rigid syntax, its EDSLs usually end up with quite a lot of syntactic noise. Thus we were able to achieve higher contrast between the original EDSL and the EDSL with a customized IDE. We used the NetBeans IDE for its native support of guarded sections, which is applied in one of the presented techniques. Its two competitors, IntelliJ IDEA and Eclipse, do not support guarding code sections out of the box -- by implementing this functionality ourselves, we would increase the costs of the IDE customizations.

\interfootnotelinepenalty=10000 

Both to demonstrate the IDE customization and to perform a controlled experiment, we used our \emph{Test-me!} language\footnote{The implementation of the language along with a usage example can be found in the following GitHub repositories: \url{http://github.com/MilanNosal/java-testDSL}, \url{https://github.com/MilanNosal/ruby-test-dsl}}. It is a DSL for defining tests for exams (inspired by the Moodle learning management system). In Listing~\ref{lst:ruby}, there is a Ruby version of a sample test. Listing~\ref{lst:java} presents a Java version of the same test (as you can see, with significantly more syntactic noise -- class definition, methods' definitions, parentheses).

\begin{listingfloat}[tbp]
\caption{An example of a biology test in \emph{Test-me!} EDSL, Ruby version}
\label{lst:ruby}
\lstset{language=Ruby,showstringspaces=false}
\begin{lstlisting}
require_relative "./TestDSL/testDSL"

create_test "Biology test", 20

multiple_choice_question "Which of the following are herbivores?", 10
incorrect_answer "Lion"
correct_answer "Sheep"
incorrect_answer "Bear"
correct_answer "Cow"

open_answer_question "What does a cat say?", 20, "Meow"

pairing_question "Combine males and females:", 10
pair "Lion", "Lioness"
pair "Bull", "Cow"
pair "Tiger", "Tigress"

run_test
\end{lstlisting}
\end{listingfloat}

\begin{listingfloat}[tbp]
\caption{An example of a biology test in \emph{Test-me!} EDSL, Java version}
\label{lst:java}
\lstset{language=Java,showstringspaces=false}
\begin{lstlisting}
package test;

import language.builder.ParsingException;
import language.builder.TestBuilder;

public class BiologyTest extends TestBuilder {
    
    @Override
    protected void define() {
        create_test("Biology test", 20);

        multiple_choice_question("Which of the following are herbivores?", 10);
        incorrect_answer("Lion");
        correct_answer("Sheep");
        incorrect_answer("Bear");
        correct_answer("Cow");

        open_answer_question("What does a cat say?", 20, "Meow");

        pairing_question("Combine males and females:", 10);
        pair("Lion", "Lioness");
        pair("Bull", "Cow");
        pair("Tiger", "Tigress");
    }
    
    public static void main(String[] args) throws ParsingException {
        new BiologyTest().compose();
    }
}
\end{lstlisting}
\end{listingfloat}

\section{IDE Reuse and Customization Techniques}

In this section, we will briefly present several techniques which reuse existing IDE features and customize them to suit the EDSL needs. As case study objects, we will use the NetBeans IDE with the Java language. However, note that the approach is not limited to this combination. Other IDEs and languages have similar possibilities. Proposed techniques include the following:

\begin{enumerate}
\item A technique for \emph{generating the boilerplate} EDSL sentence skeleton to substitute the need for a host language programmer presence during skeleton creation (section~\ref{sec:templates}).
\item A technique for \emph{hiding syntactic noise} that reduces the ``amount'' of syntactic noise presented to the EDSL user (section~\ref{sec:folding}).
\item A technique \emph{preventing inexpert editing} of the hidden syntactic noise (section~\ref{sec:guarding}).
\item A \emph{code completion support} technique that should increase user experience writing sentences in EDSL (section~\ref{sec:codeCompletion}).
\item Finally, an \emph{error reporting} technique providing clickable navigational links to domain errors (section~\ref{sec:errors}).
\end{enumerate}

We would like to emphasize that we do not claim these techniques represent the full extent into which an IDE can be inexpensively customized for the use with a pure EDSL. This is a case study with one particular IDE -- the NetBeans IDE -- and the Java language. Throughout the following sections, we also discuss the applicability of the presented techniques in a broader scope. Furthermore, different combinations of IDEs and host languages might come with other features that can be used for pure EDSL support. The goal of this work is to assess the applicability of IDE customizations for EDSLs and motivate further research in this area.


\subsection{Generating the Boilerplate}
\label{sec:templates}

In Listing~\ref{lst:java} with the test in Java, we see quite a large amount of boilerplate code: a package declaration, an import statement, a class declaration, and a method declaration along with an annotation. For a seasoned programmer, writing such code is tedious and time-consuming. For a non-programmer, such a task can be an impassable obstacle since it requires nontrivial knowledge of the implementation language.

The simplest solution to support boilerplate generation in IDEs is to use \emph{file templates}. Every IDE has built-in file templates for supported programming languages. Consider a typical Java interface creation process: A wizard asks the user for a package and interface name. Then it instantiates a file template containing a Java interface skeleton, substituting specially marked placeholders in the template with the user-supplied values. The newly created file is then opened in the IDE, ready for use.

In the same vein, we recommend creating a file template for each pure EDSL. The main idea is that templates remove the need for host language programmer involvement in the sentence skeleton creation -- the template can provide a skeleton in just a few mouse clicks. File templates are supported in all state-of-the-art IDEs, thus the technique can be used widely both in terms of IDEs (e.g., Microsoft Visual Studio and its Item Templates, file templates in JetBrains products), and host languages (.NET languages in Visual Studio, Python in PyCharm, Ruby in RubyMine, etc.).

\begin{figure}[hbt]
\centering
\includegraphics[width=\linewidth]{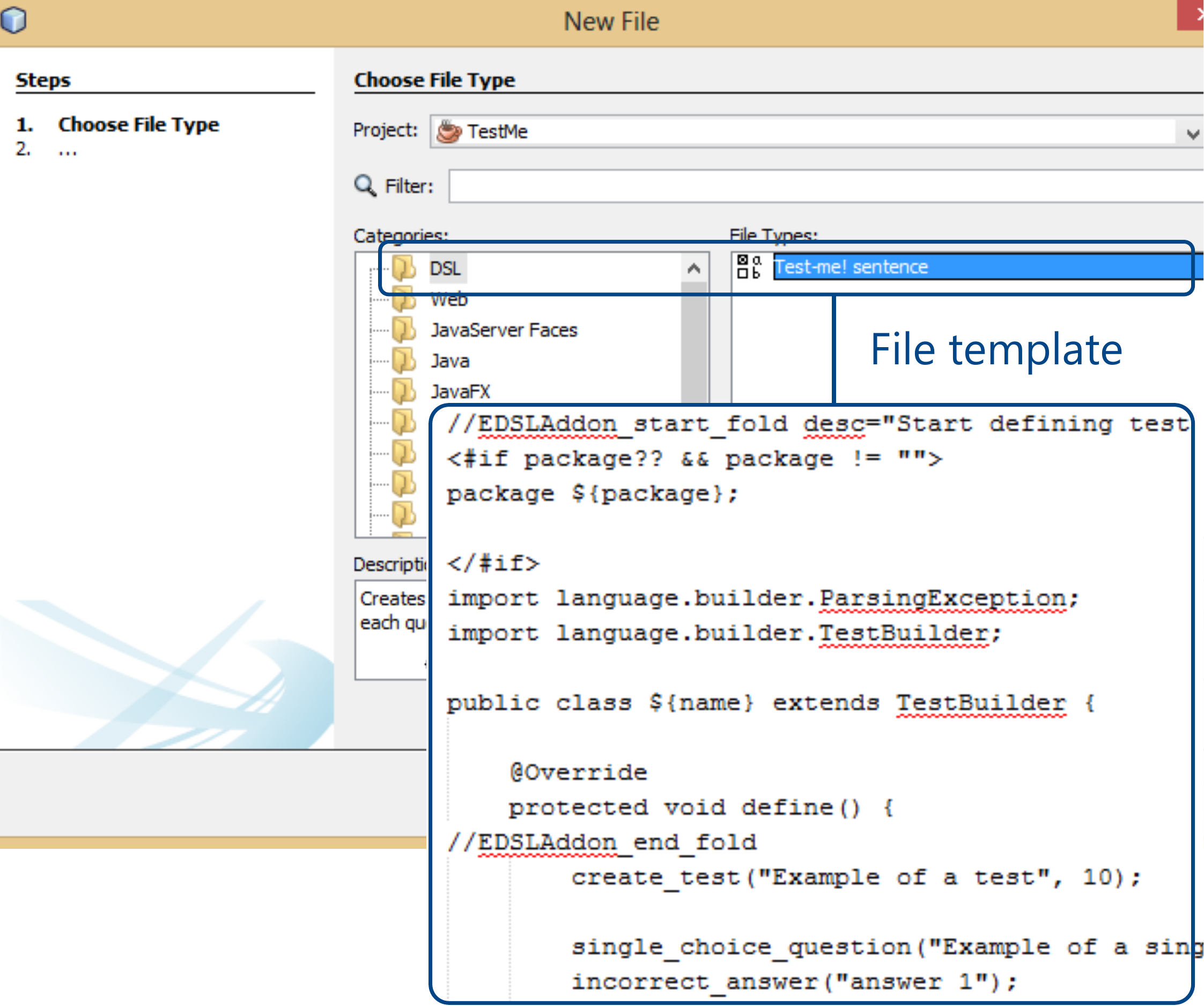}
\caption{Java-based \emph{Test-me!} EDSL sentence template}
\label{fig:templates}
\end{figure}

For our case study, we have written a NetBeans IDE template for the \emph{Test-me!} Java-based pure EDSL that creates the whole structure and an example of a test definition. The template is a part of the \emph{EDSLAddon}\footnote{\url{https://github.com/MilanNosal/edsl-addon}} plugin for the NetBeans IDE. When the plugin is installed, a user can invoke a wizard to create a new \emph{Test-me!} file. Figure~\ref{fig:templates} illustrates the \emph{Test-me!} file template.

Of course, a file template is EDSL-specific, so a single template cannot be reused for more pure EDSLs. However, for a language author, writing a template for one pure EDSL is a simple task comparable to writing a single new pure EDSL sentence.


\subsection{Hiding Syntactic Noise}
\label{sec:folding}

Creating a ready-to-edit file by a wizard offers neat support for a non-programmer trying to use a pure EDSL. However, she might still be confused by unknown, domain-unrelated concepts present in the file -- the ``syntactic noise''. Syntactic noise consists of syntactical elements of the GPL unrelated to the problem domain -- e.g., braces or import statements. For the purpose of this work, we will distinguish two types of syntactic noise depending on the relative position of its language structures to the pure EDSL sentence:

\begin{itemize}
	\item \emph{Surrounding noise} -- elements preceding or following the sentence of the pure EDSL language.
	\item \emph{Interlacing noise} -- elements that appear directly in the body of the pure EDSL sentence, but are redundant from the point of the DSL.
\end{itemize}

In the Java-based EDSL sentence introduced in Listing~\ref{lst:java}, the surrounding noise is a prefix made of the package declaration, imports, the \texttt{Biology} class declaration, its \texttt{define()} method declaration; and a postfix consisting of closing curly brackets for the class and method definitions, and the whole \texttt{main()} method. The \texttt{define()} method body is the actual pure EDSL sentence. In this case, the interlacing noise consists mainly of the parentheses and commas. Both syntactic noise types are necessary in the sentence since the code would become uncompilable by the standard Java compiler without them. Just to compare it to Ruby, in the Ruby-based EDSL sentence introduced in Listing~\ref{lst:ruby} the surrounding noise consists of the single include statement (an equivalent of the Java import section).

Our next step is to hide the syntactic noise from the user's view of the code. Ideally, we would strive to remove the syntactic noise completely. However, since we restricted ourselves to inexpensive IDE customizations, we consider removing surrounding noise good enough.

\begin{figure}[hbt]
\centering
\includegraphics[width=\linewidth]{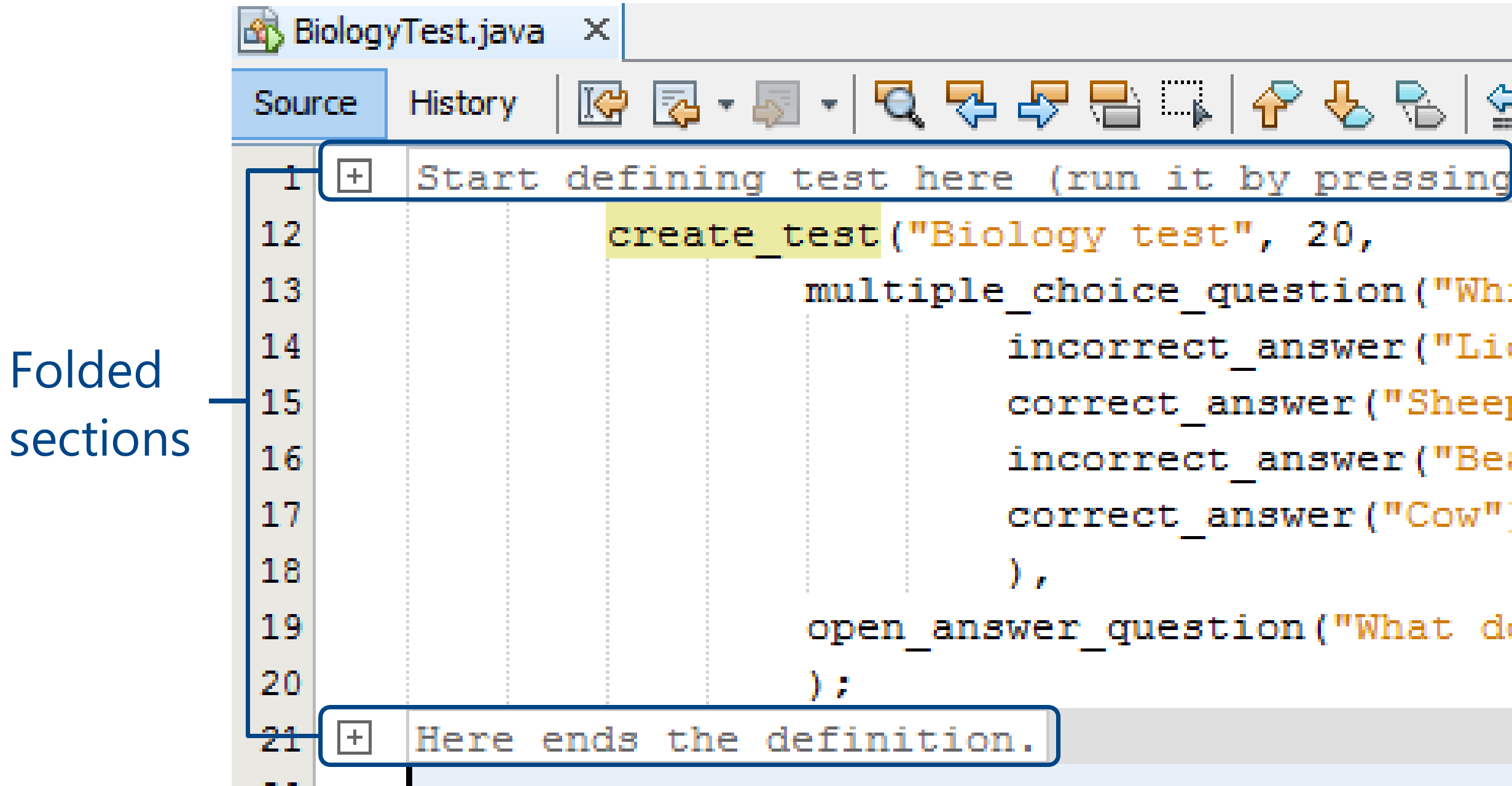}
\caption{Code folding used to hide surrounding noise}
\label{fig:folding}
\end{figure}

In IDEs, there is already a feature that allows hiding uninteresting code (uninteresting at a given time) -- \emph{code folding}. Folding some parts of the code to provide its projections is a feature found in probably all IDEs, from Xcode for Swift and Objective-C, through IntelliJ IDEA for Java, to Visual Studio for .NET languages and C++. Code folding allows collapsing (hiding) and expanding certain parts of the document via clickable icons appearing left to the source code (see Figure~\ref{fig:folding} showing folding in the NetBeans IDE). In some of the IDEs (e.g., NetBeans or IntelliJ IDEA), implementing a code fold can be as simple as adding special comments to the code. The template creator (i.e., the DSL author) can specify it using the snippet presented in Listing~\ref{lst:foldTag}.

\begin{listingfloat}[htbp]
\caption{Example of custom fold definition}
\label{lst:foldTag}
\lstset{language=Java,showstringspaces=false}
\begin{lstlisting}
// <editor-fold desc="Start defining test here" defaultstate="collapsed">
package test;
...
// </editor-fold>
\end{lstlisting}
\end{listingfloat}

In the above fragment, we defined that the fold is collapsed by default, thus hiding the syntactic noise from the perspective of the DSL user unless she explicitly expands it. Furthermore, a textual description is displayed when the fold is collapsed, giving the user hints about DSL usage (see Figure~\ref{fig:folding}).

In NetBeans IDE, specifically, we encountered a small implementation problem. We wanted to hide a method declaration, but not its body. This would cause overlapping with a built-in method body fold, which is not allowed. As a workaround, we implemented a simple NetBeans IDE plugin that re-registered the custom fold manager (a class taking care of folds) with higher priority to override the standard Java fold manager. From the implementation viewpoint, this only required a simple configuration file.

In the \emph{Test-me!} Java-based pure EDSL we were able to hide 10 significant lines of code (LOC) of the surrounding syntactic noise, and instead show 2 LOC that inform the user where the DSL sentence has to be written. If the plugin is installed, the custom tags can be used to hide practically any piece of code, for any pure EDSL (though folding of interlacing noise would be impractical and complicated due to the character of folding and the fact that interlacing noise is frequently edited).


\subsection{Preventing Inexpert Editing}
\label{sec:guarding}

Although hidden, the folded parts of the code are still sensitive to inexpert manipulation: they still can be expanded and then edited. If the declarations will get damaged, domain expert will most likely fall short of expertise to fix them, and a Java programmer will have to get involved. The solution to this issue would be an IDE feature that is able to protect code snippets of interest against any inexpert manipulation.

The NetBeans IDE provides a feature called \emph{guarded sections}\footnote{Although we claim that the techniques are not restricted to NetBeans with combination with Java, guarded sections are not featured by all the IDEs. E.g., up to the date of writing this article, two major competitors of NetBeans IDE -- IntelliJ IDEA and Eclipse -- do not provide this feature.}. A guarded section is a read-only section of a document that is protected by the editor. This way we can ensure that the hidden surrounding noise will be protected against inexpert manipulation. Guarded sections ensure the EDSL user will never invalidate sentence skeleton generated by the file template.

We added this guarding to our folds so that the section which is marked to be folded is also protected. To add guarding to folded parts, we had to implement a custom fold manager. Instead of the \texttt{<editor-fold>} XML-like tag, it uses a pair of \texttt{EDSLAddon\_start\_fold} and \texttt{EDSLAddon\_end\_fold} comments. The implementation has less than 200 LOC\footnote{See \url{https://github.com/MilanNosal/edsl-addon}} and is reusable for any pure EDSL in Java without the need of any modification.

\begin{figure}[hbt]
\centering
\includegraphics[width=\linewidth]{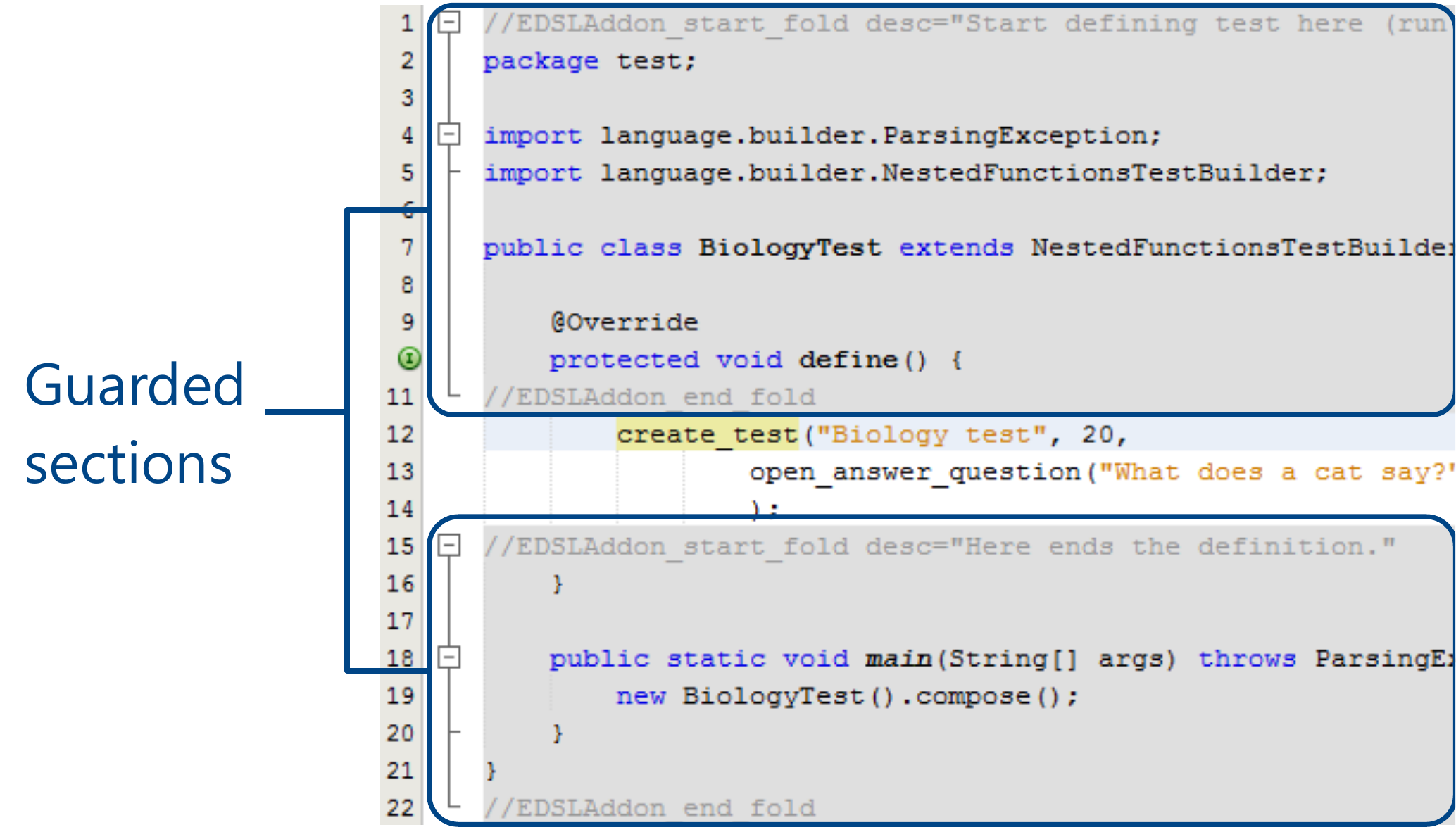}
\caption{Guarding hidden sections}
\label{fig:guarding}
\end{figure}

In Figure~\ref{fig:guarding} there is a screenshot showing code guarding in action. The NetBeans IDE highlights guarded sections with the gray background, to indicate they are disabled for editing.

When using this approach, the template creator must design the template in a way that the guarded sections will never need to be manually modified (even in case of an error). For example, in our language, all necessary classes are already imported in the imports section.


\subsection{Code Completion Support}
\label{sec:codeCompletion}

\begin{figure}[hbt]
\centering
\includegraphics[width=\linewidth]{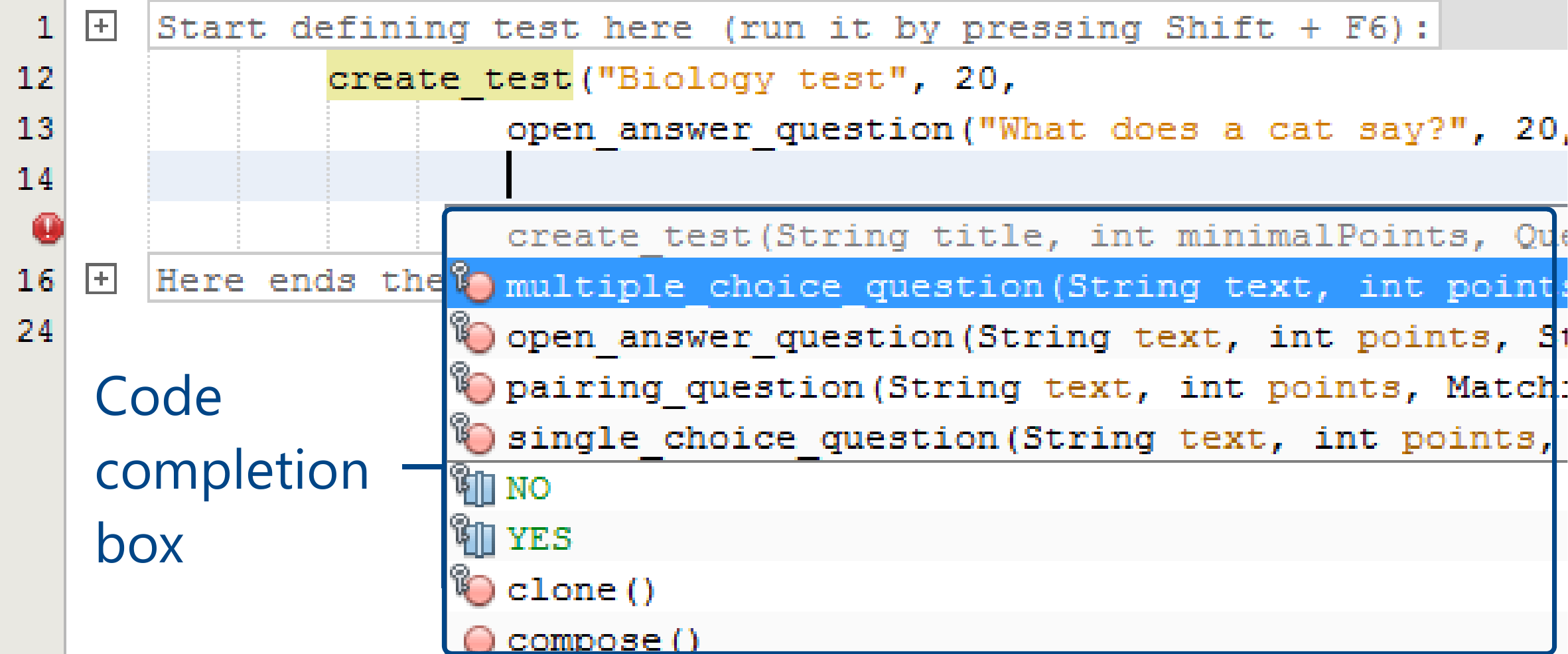}
\caption{Smart code completion suggesting domain-specific concepts}
\label{fig:completing}
\end{figure}

So far, we were interested only in syntactic noise -- the noise present in the textual representation of the DSL sentence itself. However, noise can be present also in additional IDE views (thus we will call it the \emph{view noise}) presented on top of the source code. One such view is a code completion window. When activated, it shows the developer a list of language elements from which she can choose (see Figure~\ref{fig:completing}). Often the number of suggested elements is overwhelming and the non-programming DSL user can get lost.

To cope with this problem, the NetBeans IDE\footnote{Other IDEs provide advanced code completion engines as well.} displays the most relevant suggestions above a thin gray line (see Figure~\ref{fig:completing}). As a criterion, it uses type information. For example, if a method has a parameter of type \texttt{String}, the IDE displays available string variables above the line.

When designing a DSL, the author can use this behavior with advantage. If an appropriate pure EDSL design pattern is followed, the IDE displays just relevant domain-specific concepts above the line.

In the first implementation iteration of our EDSL, we used the Nested Functions pattern\cite{fowlerDSL}, which composes functions by nesting function calls as arguments to other function calls. Nesting of methods requires that the parameters of outer method calls type-match the return types of the inner calls. This way the IDE can infer the possible method calls that can be nested in a given context.

In Figure~\ref{fig:completing} there is an example of a completion box that offers the most relevant language concepts above the thin gray line, and because of the Nested Functions and Object Scoping patterns, those are the domain-specific methods. At that place where the box is activated, the call to the \texttt{create\_test} method expects a question parameter. All proposed methods return objects of \texttt{Question} subclasses. In this process we did not have to modify the IDE, we only had to choose a suitable pure EDSL design pattern to leverage the properties of smart code completion. The Nested Functions and Object Scoping patterns can be applied for many EDSLs.

There is one additional property of code completion: it can greatly alleviate the effects of interlacing syntactic noise. Code completion generates skeletons of method calls, including quotes, comma separators, semicolons, parentheses, and other Java lexical symbols. This is another argument why IDE support is important to consider when designing a DSL.

However, after pilot testing (see sections~\ref{sec:pilot} and \ref{sec:pilot-conclusion} for a detailed discussion) we found out that even with NetBeans' code completion feature the non-programming users had serious problems working with the language. We observed that they were overwhelmed by the necessity of proper parentheses nesting -- a phenomenon that we too often see even with our first-year bachelors that are learning to program. For that reason, we decided to use the Function Sequence pattern\footnote{The Function Sequence pattern \cite{fowlerDSL} combines function calls as a sequence of statements that builds the language model. Instead of nesting function calls, they have to be only called in a specific sequence (first \texttt{create\_test()}, then \texttt{multiple\_choice\_question()}, etc.). The Ruby-based version shown in Listing~\ref{lst:ruby} also uses the Function Sequence pattern.} that was used in the Ruby version too. Listing~\ref{lst:java} presented in section~\ref{sec:introduction} already shows the final form with the Function Sequence pattern.


\subsection{Error Reporting}
\label{sec:errors}

\begin{figure}[hbt]
\centering
\includegraphics[width=\linewidth]{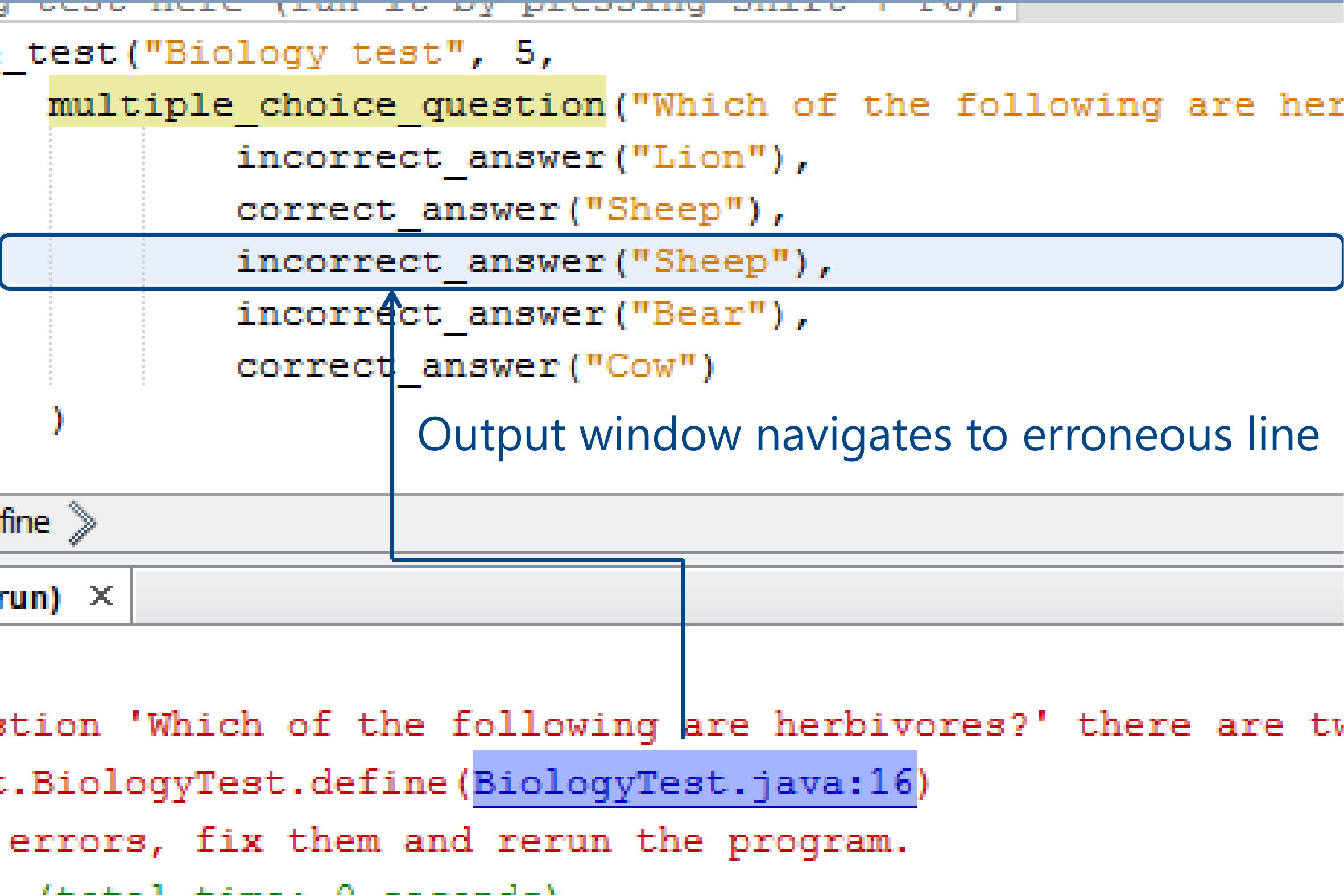}
\caption{Output windows reports a duplicated answer in a multiple choice question and navigates to it}
\label{fig:reporting}
\end{figure}

One of the challenges of DSLs is domain-specific error reporting. Suppose a user of the \emph{Test-me!} DSL mistakenly inserts the same answer into one question two times (see the top part of Figure~\ref{fig:reporting}).

A typical approach to reporting such a mistake in EDSL would be printing a simple textual message, like ``\texttt{The answer `Sheep' is a duplicate}''. An error reported this way can be very difficult to locate, especially if the DSL sentence is long, contains many occurrences of the word ``Sheep'', or if it is spread across files. Therefore it can be very useful to include also the location of the error - source file and line number. We implemented a simple utility class\footnote{\url{https://git.io/voF9k}} that allows registering model objects (e.g., of type \texttt{Answer}) and then reporting errors with links to method calls that created these objects (e.g., a specific \texttt{incorrect\_answer} call) for any pure EDSL. In our example, the error message is modified to look similar to the one in Listing~\ref{lst:errorReport}.

\begin{listingfloat}[htbp]
\caption{Error message warning about a duplicate answer with a location report}
\label{lst:errorReport}
\lstset{showstringspaces=false}
\begin{lstlisting}
The answer `Sheep' is a duplicate,
in file BiologyTest.java, line 16.
\end{lstlisting}
\end{listingfloat}

While this certainly looks better and a programmer would find the location, manually opening and scrolling the file is cumbersome. Furthermore, a non-programmer could eventually get lost when searching for the correct file in IDE windows. Therefore, we recommend providing the user with \emph{clickable navigational links} to the line in the source code that caused the domain error. To achieve this, we recommend using IDEs' output windows. IDEs usually present errors in the source code as clickable links to the error location. During printing a text to the output window, IDEs are able to recognize a pattern that defines a location in a source file, and they create a clickable link to provide easier navigation. This is common in all modern IDEs.

In Figure~\ref{fig:reporting} there is an example of an error report produced by our utility class and then presented in the NetBeans Output window. Clicking the link to the error opens the appropriate source file in the editor and puts the cursor on the given line.

Using our utility class, any Java-based pure EDSL can report domain-specific errors with clickable links in the NetBeans IDE. Generally, the approach is usable also for other languages with proper IDE support (e.g., Java with IntelliJ IDEA, or Ruby with RubyMine).

Of course, this technique does not solve the problem of pure EDSL error reporting in general. If the EDSL sentence is not a valid host sentence, then syntactic errors are still reported in terms of the host language domain. Nevertheless, we expect that providing navigational links to errors' sources can help the users to work with pure EDSLs.

\section{Experimental Evaluation}
\label{sec:experiment}

To evaluate the usefulness of the suggested IDE techniques, we conducted a controlled experiment. It consisted of a single task -- the participants tried to write a DSL sentence in the given language/IDE.  We investigated whether the IDE features used with a syntactically rigid and verbose host language can outweigh the syntactical flexibility of another language used with only a standard IDE. We compared a combination of Java-based pure EDSL and a customized NetBeans IDE (see sections~\ref{sec:templates}--\ref{sec:errors}) vs. a Ruby-based pure EDSL with a standard RubyMine IDE. The choice of the competitor was driven by the fact that Ruby is considered one of the best host languages for pure embedding \cite{ghosh2011dsl}. As an IDE we used JetBrains RubyMine since it is one of the most used Ruby IDEs \cite{rubyMine}.

\subsection{Hypotheses and Variables}

We were interested in two indicators -- effectiveness (correctness) and efficiency (correctness divided by time). Therefore, we formulate our null and alternative hypotheses as follows:

\emph{H1$_{\text{null}}$}: The proportion of participants who correctly write the test DSL sentence using Java and customized NetBeans $\leq$ the proportion of participants correctly writing it using Ruby and RubyMine.

\emph{H1$_{\text{alt}}$}: The proportion of participants who correctly write the test DSL sentence using Java and customized NetBeans $>$ the proportion of participants correctly writing it using Ruby and RubyMine.

\emph{H2$_{\text{null}}$}: The efficiency of a test DSL sentence creation with Java/NetBeans $\leq$ the efficiency with Ruby/RubyMine.

\emph{H2$_{\text{alt}}$}: The efficiency of a test DSL sentence creation with Java/NetBeans $>$ the efficiency with Ruby/RubyMine.

We will statistically test the hypotheses with a 5-percent significance level ($\alpha=5\%$)\footnote{This is an allowed probability of a Type I error, i.e., accepting $H_{null}$ when it should have been rejected instead. According to Dyb\r{a} et al. \cite{Dyba06systematic}, 0.05 is a commonly accepted norm in software engineering research.}.

There is one \emph{independent variable} -- a combination of the used host language and IDE. There are two possible values: \emph{Java} means the subjects used a Java-based pure EDSL and the NetBeans IDE customized with the presented techniques; \emph{Ruby} means they used a Ruby-based pure EDSL with a common Ruby IDE (no customization).

There are two \emph{dependent variables}. The first one is correctness of the task solution. The variable has two possible values: \emph{yes} if the user was able to correctly solve the task in the given time span, or \emph{no} if she was not.

The second dependent variable is efficiency, numerically expressed as correctness (0 or 1, since there is only one task) divided by time in minutes (with under 30 seconds precision). For a task which was not correctly completed in the allotted time frame, the efficiency was zero.

\subsection{Method}

Now we will describe the design and execution of the experiment.

\subsubsection{Participants}
\label{sec:participants}

Since our aim was to challenge quite a well-settled opinion that EDSLs are not for non-programmers, we decided to use non-programming participants. To conveniently get a larger sample of non-programming participants, we contacted a nearby high school\footnote{Gymn\'{a}zium Milana Rastislava \v{S}tef\'{a}nika v Ko\v{s}iciach, \url{http://www.gmrske.sk/}} which specialized on French foreign language (under the assumption that their students would not be programmers). Moreover, since as students they were often tested, they were familiar with the \emph{Test-me!} domain.

To test whether the potential participants fitted our non-programmer criteria, we asked each one of them to describe their previous experience with programming. Except for one\footnote{In his free time, he was writing web pages in HTML and JavaScript.} of 62 potential participants, they all claimed that they never programmed before. We queried also their Informatics teachers about the curriculum their students were going through, to ensure no threat to validity. In their curriculum, they focus on practical working with computers (using Photoshop, Microsoft Office tools, etc.), and the programming topic is presented only briefly and theoretically to provide the students with an idea what programming is. Thus, from the viewpoint of the school curriculum, they did not program.

In the end, using convenience sampling \cite{experimentation}, we obtained 62 participants. Two of them were excluded: As we already noted, one had previous programming experience. The other one was excluded because of a technical problem during the experiment execution (the experimenter made an error and provided the student with the documentation for the other version of the language, not the one the subject was supposed to work with). This provided us with a total of 60 participants. The age of the subjects spanned from 14 to 18 years.

The experiment was done during substituted lessons in the Informatics laboratory of the high school. The participation was presented as a voluntary activity -- students were given an option to reject participation without any sanction, as recommended by Carver et al. \cite{Carver03issues}. Although the participation was not awarded in any sense, the instructor tried to motivate students by presenting the materials in an interesting way and asking for their feedback during the training. At the same time, we did not fully reveal our hypotheses before the experiment \cite{Carver10checklist}. By limiting the actual experiment time to 20 minutes, we kept students focused.

\subsubsection{Design}

Each subject received only one treatment, i.e., the experiment design was unpaired. The \emph{Java} group contained 32 participants and the \emph{Ruby} group 28. The \emph{Java} group consisted of 16 students in the age of around 18 and 16 students in the age of around 14. The \emph{Ruby} group consisted of 15 students in the age of around 17 and 13 students in the age of around 15.

We assigned the participants into groups according to their attendance to the lessons which we could access. During two lessons, the students were assigned to the \emph{Java} group, students attending the other two lessons to the \emph{Ruby} group. Each lesson was attended by students of a different classroom.

\subsubsection{Materials}
\label{sec:materials}

Participants were given one of the two versions of the \emph{Test-me!} language.

The \emph{Java} group used a Java-based pure EDSL version\footnote{\url{https://github.com/MilanNosal/java-testDSL} -- the repository includes its EDSL documentation (folder ``dokumentacia'') and experiment task specification (experimentTask.pdf)} -- an example of the language sentence is presented in Listing~\ref{lst:java}. They worked with the language using NetBeans IDE with the \emph{EDSLAddon} plugin that provided an EDSL file template, folding and guarding for the \emph{Test-me!} language.

The \emph{Ruby} group used a Ruby-based version of the language\footnote{\url{https://github.com/MilanNosal/ruby-test-dsl} -- the repository includes its EDSL documentation (folder ``dokumentacia'') and experiment task specification (experimentTask.pdf)}  -- an example of the language sentence is listed in Listing~\ref{lst:ruby}. They worked with the language using the RubyMine IDE by JetBrains without any modifications.

Both implementations contained customized documentation of the language, explaining how to create a sentence in the language, documentation to each domain-specific method call, and finally instructions how to run the test. The documentation also contained a short tutorial explaining the benefits of using the IDE (code completion, error reporting). The documentation is included in the GitHub repositories of the languages (folder ``dokumentacia'').

In both cases, the language implementation generates an HTML+JavaScript document that shows the test and after filling it, the JavaScript code evaluates it. Generated HTML pages look exactly the same, regardless of the used language implementation.

All materials (and also the DSL) were localized into participants' mother tongue -- Slovak\footnote{The most recent versions of the projects on GitHub are localized to English to make it accessible to the readers of the article.}.

\subsubsection{Task}

The experiment consisted of a single task only. The participants had to implement a runnable examination test according to the specification they were given. The task specification included a screenshot of the desired output of the test definition (the generated HTML+JavaScript page for the correct solution) and a short natural language description of each question that was presented in the screenshot. The task specification is available online\footnote{\url{https://github.com/MilanNosal/java-testDSL/blob/master/experimentTask.pdf}}.

\subsubsection{Pilot Testing}
\label{sec:pilot}

Before executing the experiment, we did pilot testing. Pilot testing was done with subjects from the same high school -- but without any overlapping with the participants of the actual experiment, to prevent a serious threat to validity. After the pilot testing, we have adjusted two properties of the experiment design.

First, as we mentioned in section~\ref{sec:codeCompletion}, we finally decided to use the Function Sequence version of the Java language that does not use the code completion technique mentioned in section~\ref{sec:codeCompletion}. During pilot testing, the subjects used code completion very scarcely. However, we observed that most issues with task completion originated in difficulties with proper pairing of the parentheses in function calls. Since we used the Nested Functions pattern, proper nesting of parentheses was crucial. For this reason, we decided to remove our technique of code completion (in section~\ref{sec:codeCompletion}) from the experiment. Instead, we used the Function Sequence pattern that is also used in the Ruby version of the language (see listings~\ref{lst:java}~and~\ref{lst:ruby}). Since after this change both versions of the EDSL used the same implementation pattern, this change did not introduce a threat to validity.

Second, pilot testing determined the timing of the experiment. With each group, we had only a single 45-minute lesson which we could use for a single run of the experiment. We had to design the experiment to fit into this time. We needed around 15 minutes for introductory training and around 10 minutes for organizational tasks (preparing the environments, running the scripts to collect results, etc.). That left us with around 20 minutes for the experiment itself. Similarly, due to timing restrictions, we did not do any post-experiment satisfaction surveys with the participants.

\subsubsection{Procedure}

We executed the experiment during four substituted lessons. The first run was performed with 16 students using the Java-based language version, the second with 15 students using the Ruby-based version, third with 16 students using Java, and fourth with 13 students using Ruby again. For each experiment run, we had 45 minutes -- one teaching lesson.

Before the participants were given the task, they were given a short training to get familiar with the DSL. During the training, they were shown how to create a new source file with the necessary skeleton for writing a DSL sentence, how to define a test and a question, how to use the documentation that was provided, and finally how to run the test. During the training they were only passive observers, therefore when they were solving the task, it was the first time they were creating a sentence in the given DSL. Special attention was given to show them the benefits of using the IDE features such as code completion, inline documentation (Ruby-doc and JavaDoc), and in the \emph{Java} group also the file templates. The training took approximately 15 minutes.

We bounded the time for the task itself to 20 minutes. During the task session, the subjects could discuss task definition (if there was something unclear) with the experimenter, however, no additional discussion about the \emph{Test-me!} EDSL was allowed. In the case of questions about the EDSL, the experimenter just referred the subjects to use the language documentation.

After 20 minutes (with under 30 seconds tolerance) since the start of the task phase, the experiment was stopped. Using a script, we created snapshots of the current state of each participant's project, and videos capturing their desktops were saved.

The remaining 10 minutes of the teaching lesson was reserved for organizational tasks (preparing the environments, etc.).

\subsubsection{Data Extraction} 

The decision whether a task was fulfilled was semi-automatically performed by researchers. Automation was used to check whether the participant's solution was compilable and runnable (i.e., it did not contain any syntactic or domain-specific errors). We manually examined the tests to decide whether they exactly corresponded to the specification, or were comparably complex to the task specification\footnote{On their own request, we allowed several participants to create their own tests instead of the simple Biology test that was the goal of the task. However, to ensure unbiased results, we required them to use all the question types that were required by the task specification, to use the same number of questions with at least approximately the same number of answers. This requirement ensured they had no advantage over the rest of the participants.}. Finally, videos were skimmed to determine completion times and to observe subjects' habits during working with the DSL.

\subsection{Results}

The summary results for correctness are in the upper part of Table~\ref{tab:results}. We can clearly see that the \emph{Java} group performed better: 75\% of the subjects implemented the test correctly. In the \emph{Ruby} group, this proportion was lower (54\%).

\begin{table}
\centering
\begin{tabular}{|r|l|l|}
\hline
 & \multicolumn{2}{c|}{Group} \\ \cline{2-3}
 & Java & Ruby \\ \hline \hline
Correct & \textbf{24} (\textbf{75\%}) & \textbf{15} (\textbf{54\%}) \\ \hline
Incorrect & \textbf{8} (\textbf{25\%}) & \textbf{13} (\textbf{46\%}) \\ \hline \hline
Mean efficiency [tasks/min] & 0.044 & 0.032 \\ \hline
\end{tabular}
\caption{The summary results. Statistically significant results are in bold.}
\label{tab:results}
\end{table}

To confirm whether the results are statistically significant, we used Barnard's CSM test with confidence interval computation by Berger et al. \cite{pValues}, which is suitable for analysis of 2x2 tables. The computed p-value is 0.0406, which is less than $\alpha$ (5\%). Therefore, we reject \emph{H1$_{\text{null}}$} and accept \emph{H1$_{\text{alt}}$}. The result is statistically significant. We showed that the correctness was better for the \emph{Java} group.

The mean efficiency was also better for the Java group: 0.044 vs. 0.032 successfully completed tasks/min (see the lower part of Table~\ref{tab:results}). However, the Mann--Whitney U test on the given data produces a p-value of 0.1000 -- the result is not statistically significant.

In Table~\ref{tab:detailed}, we provide detailed experiment results for individual participants.

\begin{table}
\renewcommand{\arraystretch}{0.7}
\centering

\begin{subtable}[b]{0.45\hsize}
\centering
\begin{tabular}{|c|l|l|}
\hline
Correct & Time & Efficiency \\
  & [min] & [task/min] \\ \hline \hline
1 & 20.00 & 0.050 \\ \hline
0 & -     & 0.000 \\ \hline
1 & 20.00 & 0.050 \\ \hline
0 & -     & 0.000 \\ \hline
1 & 14.50 & 0.069 \\ \hline
1 & 14.50 & 0.069 \\ \hline
1 & 14.50 & 0.069 \\ \hline
1 & 20.00 & 0.050 \\ \hline
0 & -     & 0.000 \\ \hline
1 & 15.50 & 0.065 \\ \hline
1 & 19.00 & 0.053 \\ \hline
1 & 19.00 & 0.053 \\ \hline
1 & 20.00 & 0.050 \\ \hline
1 & 19.50 & 0.051 \\ \hline
1 & 16.50 & 0.061 \\ \hline
1 & 16.17 & 0.062 \\ \hline
1 & 16.75 & 0.060 \\ \hline
1 & 18.00 & 0.056 \\ \hline
1 & 18.50 & 0.054 \\ \hline
1 & 13.50 & 0.074 \\ \hline
0 & -     & 0.000 \\ \hline
0 & -     & 0.000 \\ \hline
0 & -     & 0.000 \\ \hline
1 & 18.25 & 0.055 \\ \hline
1 & 16.33 & 0.061 \\ \hline
0 & -     & 0.000 \\ \hline
1 & 16.33 & 0.061 \\ \hline
1 & 13.75 & 0.073 \\ \hline
0 & -     & 0.000 \\ \hline
1 & 20.00 & 0.050 \\ \hline
1 & 20.00 & 0.050 \\ \hline
1 & 14.67 & 0.068 \\ \hline
\end{tabular}
\caption{The Java group}
\label{tab:java}
\end{subtable}
\begin{subtable}[b]{0.45\hsize}
\centering
\begin{tabular}{|c|l|l|}
\hline
Correct & Time & Efficiency \\
  & [min] & [task/min] \\ \hline \hline
0 & -     & 0.000 \\ \hline
0 & -     & 0.000 \\ \hline
1 & 17.27 & 0.058 \\ \hline
0 & -     & 0.000 \\ \hline
1 & 20.00 & 0.050 \\ \hline
0 & -     & 0.000 \\ \hline
0 & -     & 0.000 \\ \hline
1 & 13.00 & 0.077 \\ \hline
1 & 19.67 & 0.051 \\ \hline
1 & 19.25 & 0.052 \\ \hline
0 & -     & 0.000 \\ \hline
1 & 15.00 & 0.067 \\ \hline
0 & -     & 0.000 \\ \hline
1 & 17.00 & 0.059 \\ \hline
1 & 12.50 & 0.080 \\ \hline
0 & -     & 0.000 \\ \hline
0 & -     & 0.000 \\ \hline
0 & -     & 0.000 \\ \hline
0 & -     & 0.000 \\ \hline
0 & -     & 0.000 \\ \hline
1 & 18.50 & 0.054 \\ \hline
1 & 13.50 & 0.074 \\ \hline
1 & 17.50 & 0.057 \\ \hline
1 & 18.58 & 0.054 \\ \hline
0 & -     & 0.000 \\ \hline
1 & 18.25 & 0.055 \\ \hline
1 & 20.00 & 0.050 \\ \hline
1 & 15.50 & 0.065 \\ \hline
\end{tabular}
\caption{The Ruby group}
\label{tab:ruby}
\end{subtable}

\caption{Experiment results for individual participants}
\label{tab:detailed}
\end{table}

\subsection{Threats to Validity}

This section discusses threats to validity of the controlled experiment according to Neto and Conte \cite{Neto14threats} and Wohlin et al. \cite{experimentation}.

\subsubsection{Construct Validity}

The chosen design of \emph{Test-me!} might be a threat to validity. As we mentioned in section~\ref{sec:materials}, after pilot testing we changed the design of the Java version of \emph{Test-me!}, which lead to significantly better results with it. This indicates that a slight change in the EDSL language syntax can significantly improve its success rate. A different design of the Ruby version could possibly end in better results with Ruby. We used the design as illustrated in Listing~\ref{lst:ruby}, because it is very simple, and it resembles natural language. Now although there is no doubt a language implementation pattern can significantly affect its success rate, this aspect should not affect the experiment because in the experiment both Java version and Ruby version used the same implementation pattern -- the Function Sequence pattern \cite{fowlerDSL}. Same implementation pattern ensures that both the control group with Java and the experimental group with Ruby had the same conditions from this aspect.

The domain selected for the experiment test should not pose a threat, since the particular domain (Biology in this case) affects only string literals of the \emph{Test-me!} EDSL, which are handled in the same way in both RubyMine and NetBeans. A test domain could only affect the motivation of the participants to finish the task -- a more interesting domain could have increased the success rate. However, since we used the same domain for both control and experimental groups, this could not affect the experiment's results.

Although the experiment consisted only of a single task to be performed by the participants, it was reasonably complex to exercise the majority of features provided by the \emph{Test-me!} DSL.

The participants were required to create an EDSL sentence from the scratch. We used this design intentionally to test also file templates, which would otherwise be useless. However, the task designed this way limits generalization of our results to creating EDSL sentences from the scratch. To be able to generalize the conclusions for program comprehension, extension and modification, we would need to perform the experiment with a task including comprehending, extending, and modifying programs. 

To measure correctness, we used just boolean values -- either correct or incorrect. It would be difficult to assess the implementations in percentages.

While the assessment of correctness was subjective (by a researcher), disputes were resolved by mutual agreement of two researchers. Since we had source code snapshots and videos available, they were resolved without time stress.

\subsubsection{Internal Validity}

Assignment to experimental groups was performed by convenience, i.e., by a classroom a student attended. Therefore, it was a quasi-experiment \cite{confounding}. However, there were no obvious confounding factors common to each classroom except age. The mean age of participants was approximately equal for both treatment groups, so this should not be a problem either. Also, the success rate of the different age groups was distributed evenly (see Table~\ref{tab:age}), which indicates that the age did not play an important role.

\begin{table}
\centering
\begin{tabular}{|l||l|l|l|l|}
\hline
Age & 14 & 15 & 17 & 18 \\ \hline
Group & Java & Ruby & Ruby & Java \\ \hline \hline
Correct & 11 & 7 & 8 & 13  \\ \hline
Incorrect & 5 & 6 & 7 & 3  \\ \hline \hline
Total & 16 & 13 & 15 & 16 \\ \hline
\end{tabular}
\caption{The results for individual age groups}
\label{tab:age}
\end{table}

Another validity threat was the quality of the training before the task. Although the same author led the training in all runs, he might have provided a worse explanation in the case of the \emph{Ruby} group. To alleviate this threat, we alternated runs, so that experience he gained would evenly affect both groups.

To mitigate instrumentation threats, the documentation mentioned in section~\ref{sec:materials} was provided to ensure they all have the same amount of information. The two versions of the documentation differed only when it was necessary -- when lexical symbols were different, and when we were talking about IDE related content. During the experiment, we observed that a common error in Ruby group was a result of copy/pasting the import section from the documentation. Higher quality of the documentation might have improved the success rate of the EDSL language. Considering that the Java version of the documentation was of equivalent quality, we can conclude that the file template customization can even alleviate shortcomings of poor documentation.

Since the experiment lasted only under 45 minutes without any prior treatment, maturation should not pose any threats to validity.

\subsubsection{External Validity}

Only students from one high school participated in our experiment, so they might not be representative of a whole population of non-programmers. However, they were familiar with the domain (from the end-user point of view) and they did not have programming experience, as it is described in section \ref{sec:participants}.

The representativeness of our EDSL, compared to real-world DSLs, is another validity threat. Although \emph{Test-me!} is not a trivial language, it has rather straightforward and simple structure (see listings~\ref{lst:java}~and~\ref{lst:ruby}). An EDSL can be much more complex from the syntactic viewpoint -- it can include variables, functions, structures, etc. We could expect that the user's benefit from our techniques (e.g., file templates) might be rendered insignificant by the complexity of the language syntax. Therefore we cannot generalize our results for all possible EDSLs. Replication of the experiment with a more complex EDSL is needed.

In this experiment, we compared just two specific combinations of a host language and IDE. The fact that we used two different IDEs designed and implemented by two different companies (RubyMine by JetBrains and NetBeans by Oracle) might be another threat to validity. Although we expect that the IDE of a dynamically typed language will be inferior to the IDE of a statically typed language, there might be other threats to validity that stem from the fact that the tools are designed by different companies. However, we need to mention that RubyMine is considered one of the best IDEs for Ruby \cite{rubyMine}, thus usage of a different IDE should not significantly affect the results.

Since our approach uses host IDEs, a learning curve of the IDE plays a role in the EDSL usage. In our experiment design, we gave the participants only a short training before the experiment. The training was without the hands-on experience. Thus the learning curve was a part of the experiment. We made this decision because the need of the whole IDE training on top of the EDSL training would be comparable to the training in the host language programming -- if we required non-programmer users to become proficient IDE users to use our EDSL, it would be the same as if we required them to learn the host language in order to use the EDSL. Therefore we have decided to include the IDE learning curve into the experiment as a part of the task. 

We have to note that we did not want to teach them in depth how all the features of IDE work -- that would be comparable to teach them how the host language works and would in effect result in clear advantage of external languages in the end. If the whole complex IDE has to be explained and understood, that puts us back at the beginning -- an external language would be a clear choice over the EDSL. However, if the IDE can be used with positive effect although the user does not understand it besides several features closely related to the EDSL (in our case our customizations), then we can consider the EDSL + IDE customizations a potential challenge to external DSLs.

\subsection{Conclusion}

In this experiment, we tried to evaluate if proper IDE support (customized NetBeans) for a pure EDSL written in an inflexible, verbose language (Java) can outweigh a succinct language with a standard IDE (Ruby and RubyMine). The experiment was performed with non-programmers (high school students), thus demonstrating that IDE support is important when non-programmers try to use a pure EDSL. Regarding correctness, the Java group performed significantly better than the Ruby group, which indicates that the customizations can outweigh the syntactic shortcomings of the host language. 

Although the experiment shows that IDE customizations can increase the non-programmers success rate with EDSLs, we cannot use its results alone to motivate the use of EDSLs over external languages in cases when non-programmers are involved. As the next step in this research, we need to compare IDE customizations with external language approaches (e.g., language workbenches), both in the aspect of user success rate and of implementation costs.

\subsubsection{Observations}

From the observation, we can conclude that the most significant reuse techniques are file templates, guarding, and folding. These techniques ensured that the non-programmer had to focus solely on the test definition. Several users of the Ruby-based EDSL failed to fulfill the task because they were not able to create an EDSL skeleton properly. In Ruby, the skeleton contained only an import statement (\texttt{require\_relative} from Listing~\ref{lst:ruby}). However, they were copy/pasting the import from the documentation, copying also a dot that ended the sentence in the documentation, and they failed to find this error and fix it. The use of a file template would prevent this issue (since no copy-pasting nor understanding of skeleton code is needed) -- thus we conclude that customization can even compensate for documentation of poor quality.

From the observation of the participants, we could conclude that the users were able to use code completion even without in-depth training of the IDE features. 10 of the total 24 successful Java users and 7 of the total 15 successful Ruby users used also code completion during solving the task. In several of those cases we noticed how they used code completion to deal with errors -- once a participant made an error and was not able to solve it, she deleted the whole erroneous line and started from the beginning using code completion. Even though they were non-programmers, they could understand the advantages of using an IDE -- providing evidence that an IDE can be really beneficial even to non-programming end users.

All the failing cases except one were due to host language syntactic errors -- confirming that syntactic noise is really the main problem of EDSLs when used by non-programmers. Only a single participant failed due to a domain error -- he used two correct answers in a single option question that did not support multiple correct answers, and he was not able to fix it in time.

We did not observe enough domain-specific errors to claim error reporting with navigational links useful. To confirm its usefulness, we could perform a new experiment focused on domain-specific errors.

Since we were capturing the screens while the participants worked, we were able to determine the approximate time of a successful solution (with under half-minute precision). In the Java group, the mean (average) time of a successful task solution was 17.3 minutes with a standard deviation of 2.3 minutes. In the Ruby group, the mean time was 17.0 minutes with a standard deviation of 2.6 minutes. This means that although the Java group performed significantly better in terms of success rate, the time was not improved when compared to the Ruby group. We believe this is a result of less interlacing syntactic noise in the case of Ruby (compare listing~\ref{lst:ruby} with listing~\ref{lst:java}) -- Ruby does not require the use of parentheses on method calls, and it also does not require a semicolon at the end of each statement. In the Java group, on the other hand, the customizations were able to help with a better success rate in solving the tasks. For example, in the Ruby group, the most often impassable problem was with the copy/pasted erroneous import statement with the dot at the ending. In the Java group, this situation could not happen -- the surrounding noise was generated using file template and prevented from editing by guarded sections, so there was no way to introduce an error there.

\subsubsection{Conclusions from Pilot Testing}
\label{sec:pilot-conclusion}

Pilot testing showed that there are pure EDSL design patterns that are very inconvenient for non-programmers. The Nested Functions pattern provides a nice fluent interface that can benefit from NetBeans smart code completion (see section~\ref{sec:codeCompletion}). However, in the case of non-programmers, the complexity of nested calls seems to by far outweigh any of their possible benefits. Observation of the pilot testing participants showed that all of them had troubles with proper parentheses nesting. Keeping the track of the open and closed parentheses is a challenging task even for our first-year students that learn to program -- non-programmers became overwhelmed by the task. In the experiment, we used the Function Sequence implementation pattern that did not nest any function calls, thus each opened parenthesis was closed before another one was opened again.

After we changed the used design pattern for our Java EDSL to the Function Sequence pattern (that was used in Ruby as well), the success rate became radically better. This indicates that the ``amount'' of syntactic noise is not so significant as the complexity of its structure (abstract syntax) -- the number of brackets stayed the same, commas were substituted by the comparable number of semicolons. The problem was the higher complexity of their structure -- brackets had to be properly nested. This observation renders the technique from section~\ref{sec:codeCompletion} unsuitable for non-programmers.

\section{Discussion}

Based on the experiment, we can see there are two different approaches that can make the IDE more useful by reducing or removing syntactic/view noise:
\begin{itemize}
	\item \emph{IDE customization} -- the language author can modify IDE's standard behavior by means of a plugin (Sections~\ref{sec:templates},~\ref{sec:folding}~and~\ref{sec:guarding}).
	\item \emph{Language manipulation} -- the language author can manipulate the pure EDSL implementation to reuse the host language IDE (Sections~\ref{sec:codeCompletion}~and~\ref{sec:errors}).
\end{itemize}

The language manipulation is non-invasive to the reused IDE infrastructure and therefore it is the less expensive alternative. The IDE modification is an invasive approach that requires more skill, but is also much more powerful. We can use the IDE API to move the language view closer to the domain while keeping its textual notation untouched.

As we could see, the user experience with pure embedding can be enhanced in multiple different aspects by IDE reuse and customization. All the techniques presented in our experiments can be easily applied to any Java-based pure EDSL without regard to its size. The language manipulation techniques required slight modifications to the language implementation, e.g., registering created objects to the utility class and then using the class to report errors. The IDE customization techniques required some effort to prepare the plugin, but the plugin itself can be reused by multiple pure EDSLs. Furthermore, the language implementation itself does not have to change. It is up to the language author to consider the options of the given IDE and to choose only techniques which are relatively cheap to use for the given pure EDSL.

We can also see there are still many issues that are inherent to pure embedding and these simple techniques cannot overcome them -- such as syntactic error reporting in DSL terms. We can also conclude from the experiment that some techniques might help programmers but are not viable for non-programming domain experts -- namely smart code completion utilization presented in section~\ref{sec:codeCompletion}.

\section{Related Work}

While there are many works contributing to the DSL field (e.g., optimization of their parsing \cite{janousek}), we focus specifically on the editing environments for DSLs. The need for proper IDE support is manifested by language workbenches that are able to provide some tool support in addition to parsers for external DSLs (for example JetBrains MPS, GME, MontiCore \cite{monti1}, Neverlang \cite{neverlang}, Rosette \cite{rosette}). There are several good overviews on language workbenches that can be used to learn more about current trends \cite{pfeiffer,workbenches,Erdweg2,mendez}. Adam et al. \cite{sorin} recognized spreadsheet-based DSLs -- DSLs consisting of data written in spreadsheets. They recognized the need of tools for these languages as well. IDE and tools support for embedding is not a novel need either, take for example the Varis tool \cite{varis} aiming to provide proper IDE support for embedded client code in PHP web applications. In our work, we focus on embedded DSLs as well; however, our primary aim is not to provide coding support for programmers using the pure EDSL, but for non-programming domain experts (while keeping additional implementation effort to a minimum).


As we have already mentioned, our work is based on the idea that when we evaluate the design of DSLs, we should consider not only their textual expressiveness but also IDE support. In the GPL context, this idea is not new. Chiba et al. \cite{Chiba2012} use IDE to introduce language syntax extension -- e.g., they implemented the \emph{Kide} editor that creates virtual source files by showing together several related pieces of code from multiple source files. Virtual source files can be used to put together code implementing crosscutting concerns, but instead of doing so on the static textual level (AspectJ), \emph{Kide} does it on the presentation level in the IDE. We used the same idea in \cite{leveraging} to aid program comprehension.

Renggli et al. \cite{renggli2010embedding} had a similar goal as our work -- they wanted to reuse the tools of the host language during the embedding. However, they do not focus on homogeneous embedded languages only, but their tool \emph{Helvetia} supports also heterogeneous embedding. Moreover, their approach requires the language author to be seasoned in context-free grammars and transformations, as \emph{Helvetia} uses an EBNF-like DSL to extend the host grammar and a DSL for transformations to modify the semantics of the homogeneous embedded DSL where needed.

Dinkelaker et al. \cite{islandGrammars} use island grammars to specify concrete syntax of an embedded DSL. They use annotations on program elements to define concrete syntax of a previously homogeneous embedded DSL and, in turn, they use the IDE plugin to generate a corresponding preprocessor and language support such as code completion. This changes the language to heterogeneous with the implementation costs lower than in other approaches (like for example Metaborg \cite{metaborg}). Kur\v{s} et al. \cite{boundedSeas} use island grammars for creating composable and reusable islands called \emph{bounded seas} that compute scope for water parsing. We aim for the same result as these approaches, but we keep the language homogeneously embedded.

Scherr et al. \cite{scherr} present a prototype for staging shallow embedded DSLs to get benefits of deep embedding. Both shallow and deep embedding concerns homogeneous embedded DSLs. In our work we used shallow embedding -- the calls were directly interpreted and the HTML test was generated. Deep embedding creates an intermediate representation (IR) of the language sentence that allows further processing -- optimization, etc. Staging is the process of creating an IR for shallow embedded DSLs. While the prototype created by Scherr et al. \cite{scherr} is based on Java annotations, they recommend solidifying embedding as a first-class feature of a host GPL (e.g., by using keywords for staging instead of their custom annotations). Native support for embedding in a GPL could positively affect IDE reusability for embedded DSLs. They do not further address the problem of IDE reuse for domain experts. Quite a similar approach is proposed by Erdweg et al. \cite{growing}, who basically focus on native support for embedding directly at the IDE level. They propose to organize editor features into editor libraries that should provide a simple way to extend the IDE to support the embedded language.

Considering that our approach projects the pure embedded DSL sentence into a different presentation syntax (it hides surrounding noise), we could relate it to projectional editors \cite{voelter}. Usually, the goal of projectional editing is to provide a more flexible environment for language composition and extension. We used a similar idea to provide the non-programming domain expert better syntax for working with the language.

\section{Conclusions}

In this paper, we experimented with the host IDE reuse and customization to involve non-programmers in using a pure EDSL. We presented several techniques and recommendations that aim to support non-programmers' involvement. File templates can be used to generate the code skeleton (surrounding noise), so that the user does not have to write it herself (which might be an impassable obstacle for a non-programmer). Folding and guarding aimed to reduce syntactic noise by hiding and protecting the surrounding noise from the language user. Additionally, we proposed to use the Nested Functions pattern to utilize smart code completion so that it would prefer domain-specific method calls instead of other host language constructs. Finally, we recommended using error messages with navigational links to increase efficiency in working with pure EDSLs. While the techniques were presented using a case study based on the NetBeans IDE, most features reused by the proposed techniques exist in other IDEs as well, thus we assume wider applicability.

We conducted a controlled experiment to evaluate the proposed techniques. The experiment compared non-programmers' work with Ruby and non-modified RubyMine IDE versus non-programmers' work with Java and NetBeans IDE customized to use the proposed techniques. The results confirmed that proposed techniques can even outweigh the inherent issues with inflexible syntax of the host language. On the other hand, the experiment indicated that the smart code completion utilization as proposed in section~\ref{sec:codeCompletion} brings more problems than benefits for the non-programmers. Of course, drawing conclusions from one experiment is not as certain as from a family of experiments \cite{Kosar12program}, so replications by other researchers are welcome.

To sum up the article, when we evaluate the design of DSLs, we should consider not only their textual expressiveness but also IDE support, since it can improve the user experience of even non-programming domain experts while keeping creation costs relatively low (the main advantage of the EDSL implementation approach). 

There are multiple directions for possible future work. The following two options are our priority. First, we need to analyze the applicability of our recommendations and techniques in different combinations of a host language and IDE (e.g., Java with IntelliJ IDEA, Ruby with RubyMine, etc.), since our experiment involved only Java with NetBeans IDE. Next, we would like to compare proposed techniques with stand-alone (external) DSLs. If a pure EDSL could compete with a stand-alone DSL just on the basis of IDE reuse, it could contribute to solving the open problem of selecting the viable implementation strategy for DSLs \cite{Bagge}.

One of the EDSL implementation strategies are source code annotations. Annotations along with their processor form a formal language, and they provide a framework for language composition \cite{nasComsis}. Source code annotations are one of the most commonly used formats for configuration languages \cite{mojComsis}, with many different usage domains (for example, we used annotations to annotate the source code with its intent \cite{nasComlan}). In the future work, we would like to explore the opportunities to support annotation-based EDSLs using the host IDE.

\section*{Acknowledgment}

We would like to give our thanks to the director and the teachers of Gymn\'{a}zium Milana Rastislava \v{S}tef\'{a}nika v Ko\v{s}iciach (\url{http://www.gmrske.sk/}) for giving us the opportunity to perform the experiment at their school. The thanks also belong to all the students that voluntarily participated in the experiment.

This work was supported by project KEGA No. 019TUKE-4/2014 Integration of the Basic Theories of Software Engineering into Courses for Informatics Master Study Programmes at Technical Universities -- Proposal and Implementation.

\section*{References}
\bibliography{ide-reuse}

\begin{thebibliography}{10}
\expandafter\ifx\csname url\endcsname\relax
  \def\url#1{\texttt{#1}}\fi
\expandafter\ifx\csname urlprefix\endcsname\relax\def\urlprefix{URL }\fi
\expandafter\ifx\csname href\endcsname\relax
  \def\href#1#2{#2} \def\path#1{#1}\fi

\bibitem{preliminaryStudy}
T.~Kosar, P.~E. Mart\'{i}nez~L\'{o}pez, P.~A. Barrientos, M.~Mernik, {A
  preliminary study on various implementation approaches of domain-specific
  language}, Information and Software Technology 50~(5) (2008) 390--405.
\newblock \href {http://dx.doi.org/10.1016/j.infsof.2007.04.002}
  {\path{doi:10.1016/j.infsof.2007.04.002}}.

\bibitem{whenAndHow}
M.~Mernik, J.~Heering, A.~M. Sloane, {When and How to Develop Domain-specific
  Languages}, ACM Computing Surveys 37~(4) (2005) 316--344.
\newblock \href {http://dx.doi.org/10.1145/1118890.1118892}
  {\path{doi:10.1145/1118890.1118892}}.

\bibitem{dslVanDeursen}
A.~van Deursen, P.~Klint, J.~Visser, {Domain-specific languages: an annotated
  bibliography}, SIGPLAN Notices 35~(6) (2000) 26--36.
\newblock \href {http://dx.doi.org/10.1145/352029.352035}
  {\path{doi:10.1145/352029.352035}}.

\bibitem{designPatternsDSL}
D.~Spinellis, {Notable design patterns for domain-specific languages}, Journal
  of Systems and Software 56~(1) (2001) 91--99.
\newblock \href {http://dx.doi.org/10.1016/S0164-1212(00)00089-3}
  {\path{doi:10.1016/S0164-1212(00)00089-3}}.

\bibitem{hudak1997domain}
P.~Hudak, Domain-specific languages, Handbook of Programming Languages 3 (1997)
  39--60.

\bibitem{mernikFAL}
S.~Zawoad, M.~Mernik, R.~Hasan, {Towards Building a Forensics Aware Language
  for Secure Logging}, Computer Science and Information Systems 11~(4) (2014)
  1291--1314.
\newblock \href {http://dx.doi.org/10.2298/CSIS131201051Z}
  {\path{doi:10.2298/CSIS131201051Z}}.

\bibitem{questionnaireDSL}
M.~Gouseti, C.~Peters, T.~van~der Storm, Extensible language implementation
  with object algebras (short paper), ACM SIGPLAN Notices 50~(3) (2015) 25--28.
\newblock \href {http://dx.doi.org/10.1145/2775053.2658765}
  {\path{doi:10.1145/2775053.2658765}}.

\bibitem{sonjaConstraint}
S.~Risti{\'c}, S.~Aleksi{\'c}, M.~{\v{C}}elikovi{\'c}, I.~Lukovi{\'c}, {Generic
  and Standard Database Constraint Meta-Models}, Computer Science and
  Information Systems 11~(2) (2014) 679--696.
\newblock \href {http://dx.doi.org/10.2298/CSIS140216037R}
  {\path{doi:10.2298/CSIS140216037R}}.

\bibitem{vranicOrgPatterns}
T.~Fr\v{t}ala, V.~Vrani\'{c}, {Animating Organizational Patterns}, in:
  Proceedings of the Eighth International Workshop on Cooperative and Human
  Aspects of Software Engineering, CHASE '15, IEEE Press, 2015, pp. 8--14.
\newblock \href {http://dx.doi.org/10.1109/CHASE.2015.8}
  {\path{doi:10.1109/CHASE.2015.8}}.

\bibitem{polap}
M.~Wo{\'z}niak, D.~Po{\l}ap, C.~Napoli, E.~Tramontana, Graphic object feature
  extraction system based on {C}uckoo search algorithm, Expert Systems with
  Applications 66~(C) (2016) 20--31.
\newblock \href {http://dx.doi.org/10.1016/j.eswa.2016.08.068}
  {\path{doi:10.1016/j.eswa.2016.08.068}}.

\bibitem{extensibleLanguage}
S.~Chodarev, J.~Koll\'{a}r, Extensible host language for domain-specific
  languages, Computing \& Informatics 35~(1) (2016) 84--110.

\bibitem{mernik2013}
M.~Mernik, {An object-oriented approach to language compositions for software
  language engineering}, Journal of Systems and Software 86~(9) (2013)
  2451--2464.
\newblock \href {http://dx.doi.org/10.1016/j.jss.2013.04.087}
  {\path{doi:10.1016/j.jss.2013.04.087}}.

\bibitem{infoQ}
InfoQ, {Developing a Complex External DSL},
  \url{https://www.infoq.com/articles/External-DSL-Vaughn-Vernon}, accessed:
  2016-11-20.

\bibitem{fowlerDSL}
M.~Fowler, {Domain-Specific Languages}, 1st Edition, Addison-Wesley
  Professional, 2010.

\bibitem{erikM}
W.~R. Cook, {Erik Meijer started a discussion on Domain Specific Languages},
  \url{http://lambda-the-ultimate.org/node/4560}, accessed: 2016-11-20.

\bibitem{boundedSeas}
J.~Kur{\v{s}}, M.~Lungu, R.~Iyadurai, O.~Nierstrasz, {Bounded seas}, Computer
  Languages, Systems \& Structures 44, Part A (2015) 114--140.
\newblock \href {http://dx.doi.org/10.1016/j.cl.2015.08.002}
  {\path{doi:10.1016/j.cl.2015.08.002}}.

\bibitem{hudak}
P.~Hudak, {Building domain-specific embedded languages}, ACM Computing Surveys
  28~(4es) (1996) 1--6.
\newblock \href {http://dx.doi.org/10.1145/242224.242477}
  {\path{doi:10.1145/242224.242477}}.

\bibitem{cunningham}
H.~C. Cunningham, {A little language for surveys: constructing an internal DSL
  in Ruby}, in: Proceedings of the 46th ACM Southeast Conference, ACM-SE 46,
  ACM, New York, NY, USA, 2008, pp. 282--287.
\newblock \href {http://dx.doi.org/10.1145/1593105.1593181}
  {\path{doi:10.1145/1593105.1593181}}.

\bibitem{embeddingRuby}
T.~Grust, M.~Mayr, {A Deep Embedding of Queries into Ruby}, in: IEEE 28th
  International Conference on Data Engineering (ICDE), 2012, pp. 1257--1260.
\newblock \href {http://dx.doi.org/10.1109/ICDE.2012.121}
  {\path{doi:10.1109/ICDE.2012.121}}.

\bibitem{freeman}
S.~Freeman, N.~Pryce, {Evolving an embedded domain-specific language in Java},
  in: Companion to the 21st ACM SIGPLAN symposium on Object-oriented
  programming systems, languages, and applications, OOPSLA '06, ACM, New York,
  NY, USA, 2006, pp. 855--865.
\newblock \href {http://dx.doi.org/10.1145/1176617.1176735}
  {\path{doi:10.1145/1176617.1176735}}.

\bibitem{ghosh2011dsl}
D.~Ghosh, {DSL} for the uninitiated, Communications of the ACM 54~(7) (2011)
  44--50.
\newblock \href {http://dx.doi.org/10.1145/1965724.1965740}
  {\path{doi:10.1145/1965724.1965740}}.

\bibitem{DSLSystematicMappingStudy}
L.~M. do~Nascimento, D.~L. Viana, P.~A.~S. Neto, D.~A. Martins, V.~C. Garcia,
  S.~R. Meira, {A Systematic Mapping Study on Domain-specific Languages}, in:
  Proceedings of the 7th International Conference on Software Engineering
  Advances, ICSEA'12, 2012, pp. 179--187.

\bibitem{SMSKosar}
T.~Kosar, S.~Bohra, M.~Mernik, {Domain-Specific Languages: A Systematic Mapping
  Study}, Information and Software Technology 71~(C) (2016) 77--91.
\newblock \href {http://dx.doi.org/10.1016/j.infsof.2015.11.001}
  {\path{doi:10.1016/j.infsof.2015.11.001}}.

\bibitem{rubyMine}
StackOverflow, {What Ruby IDE do you prefer?},
  \url{http://stackoverflow.com/questions/16991/what-ruby-ide-do-you-prefer},
  accessed: 2016-06-30.

\bibitem{Dyba06systematic}
T.~Dyb\r{a}, V.~B. Kampenes, D.~I. Sj{\o}berg, A systematic review of
  statistical power in software engineering experiments, Information and
  Software Technology 48~(8) (2006) 745 -- 755.
\newblock \href {http://dx.doi.org/10.1016/j.infsof.2005.08.009}
  {\path{doi:10.1016/j.infsof.2005.08.009}}.

\bibitem{experimentation}
C.~Wohlin, P.~Runeson, M.~H\"ost, M.~C. Ohlsson, B.~Regnell, A.~Wessl\'en,
  {Experimentation in Software Engineering}, Springer Publishing Company,
  Incorporated, 2012.

\bibitem{Carver03issues}
J.~Carver, L.~Jaccheri, S.~Morasca, F.~Shull, Issues in using students in
  empirical studies in software engineering education, in: Proceedings of the
  9th International Symposium on Software Metrics, METRICS '03, IEEE Computer
  Society, Washington, DC, USA, 2003, pp. 239--249.
\newblock \href {http://dx.doi.org/10.1109/METRIC.2003.1232471}
  {\path{doi:10.1109/METRIC.2003.1232471}}.

\bibitem{Carver10checklist}
J.~C. Carver, L.~Jaccheri, S.~Morasca, F.~Shull, A checklist for integrating
  student empirical studies with research and teaching goals, Empirical
  Software Engineering 15~(1) (2010) 35--59.
\newblock \href {http://dx.doi.org/10.1007/s10664-009-9109-9}
  {\path{doi:10.1007/s10664-009-9109-9}}.

\bibitem{pValues}
R.~L. Berger, D.~D. Boos, {P Values Maximized Over a Confidence Set for the
  Nuisance Parameter}, Journal of the American Statistical Association 89~(427)
  (1994) 1012--1016.
\newblock \href {http://dx.doi.org/10.1080/01621459.1994.10476836}
  {\path{doi:10.1080/01621459.1994.10476836}}.

\bibitem{Neto14threats}
A.~A. Neto, T.~Conte, Threats to validity and their control actions -- results
  of a systematic literature review, Technical Report TR-USES-2014-0002,
  Universidade Federal do Amazonas (Mar. 2014).

\bibitem{confounding}
J.~Siegmund, J.~Schumann, {Confounding parameters on program comprehension: a
  literature survey}, Empirical Software Engineering 20~(4) (2015) 1159--1192.
\newblock \href {http://dx.doi.org/10.1007/s10664-014-9318-8}
  {\path{doi:10.1007/s10664-014-9318-8}}.

\bibitem{janousek}
R.~Pol\'{a}ch, J.~Tr\'{a}vn\'{i}\v{c}ek, J.~Janou\v{s}ek, B.~Melichar,
  {Efficient determinization of visibly and height-deterministic pushdown
  automata}, Computer Languages, Systems \& Structures 46 (2016) 91--105.
\newblock \href {http://dx.doi.org/10.1016/j.cl.2016.07.005}
  {\path{doi:10.1016/j.cl.2016.07.005}}.

\bibitem{monti1}
H.~Gr\"{o}nniger, H.~Krahn, B.~Rumpe, M.~Schindler, S.~V\"{o}lkel, {MontiCore:
  a framework for the development of textual domain specific languages}, in:
  Companion of the 30th international conference on Software engineering, ICSE
  Companion '08, ACM, New York, NY, USA, 2008, pp. 925--926.
\newblock \href {http://dx.doi.org/10.1145/1370175.1370190}
  {\path{doi:10.1145/1370175.1370190}}.

\bibitem{neverlang}
E.~Vacchi, W.~Cazzola, {Neverlang: {A} framework for feature-oriented language
  development}, Computer Languages, Systems {\&} Structures 43 (2015) 1--40.
\newblock \href {http://dx.doi.org/10.1016/j.cl.2015.02.001}
  {\path{doi:10.1016/j.cl.2015.02.001}}.

\bibitem{rosette}
E.~Torlak, R.~Bodik, {Growing Solver-aided Languages with Rosette}, in:
  Proceedings of the 2013 ACM International Symposium on New Ideas, New
  Paradigms, and Reflections on Programming \& Software, Onward! 2013, ACM, New
  York, NY, USA, 2013, pp. 135--152.
\newblock \href {http://dx.doi.org/10.1145/2509578.2509586}
  {\path{doi:10.1145/2509578.2509586}}.

\bibitem{pfeiffer}
M.~Pfeiffer, J.~Pichler, {A Comparison of Tool Support for Textual
  Domain-Specific Languages}, in: Proceedings of the 8th OOPSLA Workshop on
  Domain-Specific Modeling, 2008, pp. 1--7.

\bibitem{workbenches}
S.~Erdweg, T.~van~der Storm, M.~V\"{o}lter, M.~Boersma, R.~Bosman, W.~Cook,
  A.~Gerritsen, A.~Hulshout, S.~Kelly, A.~Loh, G.~D. Konat, P.~J. Molina,
  M.~Palatnik, R.~Pohjonen, E.~Schindler, K.~Schindler, R.~Solmi, V.~A. Vergu,
  E.~Visser, K.~van~der Vlist, G.~H. Wachsmuth, J.~van~der Woning, {The State
  of the Art in Language Workbenches}, in: Software Language Engineering, Vol.
  8225 of Lecture Notes in Computer Science, Springer International Publishing,
  2013, pp. 197--217.
\newblock \href {http://dx.doi.org/10.1007/978-3-319-02654-1_11}
  {\path{doi:10.1007/978-3-319-02654-1_11}}.

\bibitem{Erdweg2}
S.~Erdweg, T.~van~der Storm, M.~V\"{o}lter, L.~Tratt, R.~Bosman, W.~R. Cook,
  A.~Gerritsen, A.~Hulshout, S.~Kelly, A.~Loh, G.~Konat, P.~J. Molina,
  M.~Palatnik, R.~Pohjonen, E.~Schindler, K.~Schindler, R.~Solmi, V.~Vergu,
  E.~Visser, K.~van~der Vlist, G.~Wachsmuth, J.~van~der Woning, {Evaluating and
  comparing language workbenches: Existing results and benchmarks for the
  future}, Computer Languages, Systems \& Structures 44, Part A (2015) 24--47.
\newblock \href {http://dx.doi.org/10.1016/j.cl.2015.08.007}
  {\path{doi:10.1016/j.cl.2015.08.007}}.

\bibitem{mendez}
D.~M\'{e}ndez-Acuna, J.~A. Galindo, T.~Degueule, B.~Combemale, B.~Baudry,
  {Leveraging Software Product Lines Engineering in the development of external
  DSLs: A systematic literature review}, Computer Languages, Systems \&
  Structures 46 (2016) 206--235.
\newblock \href {http://dx.doi.org/10.1016/j.cl.2016.09.004}
  {\path{doi:10.1016/j.cl.2016.09.004}}.

\bibitem{sorin}
S.~Adam, U.~P. Schultz, {Towards Tool Support for Spreadsheet-based
  Domain-specific Languages}, SIGPLAN Notices 51~(3) (2015) 95--98.
\newblock \href {http://dx.doi.org/10.1145/2936314.2814215}
  {\path{doi:10.1145/2936314.2814215}}.

\bibitem{varis}
H.~V. Nguyen, C.~K\"{a}stner, T.~N. Nguyen, {Varis: IDE Support for Embedded
  Client Code in PHP Web Applications}, in: Proceedings of the 37th
  International Conference on Software Engineering - Volume 2, ICSE '15, IEEE
  Press, Piscataway, NJ, USA, 2015, pp. 693--696.
\newblock \href {http://dx.doi.org/10.1109/ICSE.2015.225}
  {\path{doi:10.1109/ICSE.2015.225}}.

\bibitem{Chiba2012}
S.~Chiba, M.~Horie, K.~Kanazawa, F.~Takeyama, Y.~Teramoto, {Do We Really Need
  to Extend Syntax for Advanced Modularity?}, in: Proceedings of the 11th
  Annual International Conference on Aspect-oriented Software Development, AOSD
  '12, ACM, New York, NY, USA, 2012, pp. 95--106.
\newblock \href {http://dx.doi.org/10.1145/2162049.2162061}
  {\path{doi:10.1145/2162049.2162061}}.

\bibitem{leveraging}
J.~Porub{\"a}n, M.~Nos{\'a}\v{l}, {Leveraging Program Comprehension with
  Concern-oriented Source Code Projections}, in: 3rd Symposium on Languages,
  Applications and Technologies, Vol.~38 of OpenAccess Series in Informatics
  (OASIcs), Schloss Dagstuhl--Leibniz-Zentrum fuer Informatik, Dagstuhl,
  Germany, 2014, pp. 35--50.
\newblock \href {http://dx.doi.org/10.4230/OASIcs.SLATE.2014.35}
  {\path{doi:10.4230/OASIcs.SLATE.2014.35}}.

\bibitem{renggli2010embedding}
L.~Renggli, T.~G{\^\i}rba, O.~Nierstrasz, {Embedding languages without breaking
  tools}, in: Proceedings of the 24th European conference on Object-oriented
  programming, ECOOP'10, Springer, 2010, pp. 380--404.
\newblock \href {http://dx.doi.org/10.1007/978-3-642-14107-2_19}
  {\path{doi:10.1007/978-3-642-14107-2_19}}.

\bibitem{islandGrammars}
T.~Dinkelaker, M.~Eichberg, M.~Mezini, {Incremental concrete syntax for
  embedded languages with support for separate compilation}, Science of
  Computer Programming 78~(6) (2013) 615--632.
\newblock \href {http://dx.doi.org/10.1016/j.scico.2012.12.002}
  {\path{doi:10.1016/j.scico.2012.12.002}}.

\bibitem{metaborg}
M.~Bravenboer, R.~de~Groot, E.~Visser, {MetaBorg in action: examples of
  domain-specific language embedding and assimilation using Stratego/XT}, in:
  Proceedings of the 2005 international conference on Generative and
  Transformational Techniques in Software Engineering, GTTSE'05,
  Springer-Verlag, Berlin, Heidelberg, 2006, pp. 297--311.
\newblock \href {http://dx.doi.org/10.1007/11877028_10}
  {\path{doi:10.1007/11877028_10}}.

\bibitem{scherr}
M.~Scherr, S.~Chiba, {Almost First-class Language Embedding: Taming Staged
  Embedded DSLs}, SIGPLAN Notices 51~(3) (2015) 21--30.
\newblock \href {http://dx.doi.org/10.1145/2936314.2814217}
  {\path{doi:10.1145/2936314.2814217}}.

\bibitem{growing}
S.~Erdweg, L.~C. Kats, T.~Rendel, C.~K\"{a}stner, K.~Ostermann, E.~Visser,
  {Growing a Language Environment with Editor Libraries}, SIGPLAN Notices
  47~(3) (2011) 167--176.
\newblock \href {http://dx.doi.org/10.1145/2189751.2047891}
  {\path{doi:10.1145/2189751.2047891}}.

\bibitem{voelter}
M.~Voelter, J.~Siegmund, T.~Berger, B.~Kolb, {Towards user-friendly
  projectional editors}, in: 7th International Conference on Software Language
  Engineering, SLE 2014, Springer, 2014, pp. 41--61.
\newblock \href {http://dx.doi.org/10.1007/978-3-319-11245-9_3}
  {\path{doi:10.1007/978-3-319-11245-9_3}}.

\bibitem{Kosar12program}
T.~Kosar, M.~Mernik, J.~C. Carver, Program comprehension of domain-specific and
  general-purpose languages: Comparison using a family of experiments,
  Empirical Software Engineering 17~(3) (2012) 276--304.
\newblock \href {http://dx.doi.org/10.1007/s10664-011-9172-x}
  {\path{doi:10.1007/s10664-011-9172-x}}.

\bibitem{Bagge}
A.~H. Bagge, V.~Zaytsev, {Open and Original Problems in Software Language
  Engineering 2015 Workshop Report}, SIGSOFT Software Engineering Notes 40~(3)
  (2015) 32--37.
\newblock \href {http://dx.doi.org/10.1145/2757308.2757313}
  {\path{doi:10.1145/2757308.2757313}}.

\bibitem{nasComsis}
M.~Nos\'{a}\v{l}, M.~Sul\'{i}r, J.~Juh\'{a}r, {Language Composition Using
  Source Code Annotations}, Computer Science and Information Systems 13 (2016)
  707--729.
\newblock \href {http://dx.doi.org/10.2298/CSIS160114024N}
  {\path{doi:10.2298/CSIS160114024N}}.

\bibitem{mojComsis}
M.~Nos\'{a}\v{l}, J.~Porub\"{a}n, {XML to Annotations Mapping Definition with
  Patterns}, Computer Science and Information Systems 11 (2014) 1455--1477.
\newblock \href {http://dx.doi.org/10.2298/CSIS130920049N}
  {\path{doi:10.2298/CSIS130920049N}}.

\bibitem{nasComlan}
M.~Sul\'{\i}r, M.~Nos\'{a}\v{l}, J.~Porub\"{a}n, {Recording Concerns in Source
  Code Using Annotations}, Computer Languages, Systems \& Structures 46~(C)
  (2016) 44--65.
\newblock \href {http://dx.doi.org/10.1016/j.cl.2016.07.003}
  {\path{doi:10.1016/j.cl.2016.07.003}}.

\end{thebibliography}

\section*{Vitae}

\textbf{Milan Nos\'{a}\v{l}} is currently working as a full-time iOS developer at Svagant, Ko\v{s}ice. He received his PhD. in Computer Science in 2015 for his work ``Leveraging Program Comprehension with Concern-oriented Projections'' from Technical university of Ko\v{s}ice, Slovakia. In his free time, he works on research in the field of attribute-oriented programming (source code annotations), program comprehension, projectional editors, and domain-specific languages.

\textbf{Jaroslav Porub\"{a}n} is an Associate professor and the Head of Department of Computers and Informatics, Technical university of Ko\v{s}ice, Slovakia. He received his MSc. in Computer Science in 2000 and his PhD. in Computer Science in 2004. Since 2003 he is a member of the Department of Computers and Informatics at Technical University of Ko\v{s}ice. Currently, the main subject of his research is the computer language engineering concentrating on design and implementation of domain specific languages and computer language composition and evolution.

\textbf{Mat\'{u}\v{s} Sul\'{i}r} is a PhD student at the Department of Computers and Informatics, Faculty of Electrical Engineering and Informatics, Technical University of Ko\v{s}ice. He graduated with a master's degree in Computer Science in 2014. His current research is focused on program comprehension, source code annotations, and empirical methods in software engineering.

\listoftodos

\end{document}